\documentclass{xappolb}
\usepackage{epsfig}

\begin{document}
\title{ISI/FSI in Threshold Meson Production --- Onshell
Approach and Coulomb Problem%
\thanks{Proceedings of MESON 2000, Cracow, Poland, May 19--23, 2000, to be published in Acta Physica Polonica {\bf B}}%
}
\author{Frieder Kleefeld
\address{Centro de F\'{\i}sica das Interac\c{c}\~{o}es Fundamentais, 
Instituto Superior T\'{e}cnico, Edif\'{\i}cio Ci\^{e}ncia, Piso 3,  
Av. Rovisco Pais, 
P-1049-001 LISBOA,
Portugal\\
 e-mail: {\tt kleefeld@gtae3.ist.utl.pt}}
}
\maketitle
\begin{abstract}
The onshell description of Initial-- and Final--State--Interactions (ISI and FSI) of threshold meson production reactions  is reviewed. Existing  onshell models and their offshell extension are discussed. Unitarity constraints on enhancement factors are formulated. A strategy for the treatment of essential singularities connected to Coulomb--like FSI is given.
\end{abstract}
\ \\  
\par ISI and FSI strongly determine qualitatively and quantitatively the energy dependence of total cross sections of particle production processes close to threshold. Following the Watson--Migdal approach \cite{wat1,kle1,kle2} it is assumed that the T--matrix $T_{fi}$ can be ``factorised'' into a product of a short ranged production amplitude $T^{\,(0)}_{fi}$ and ``enhancement factors'' $T_{{}_{ISI}} (i)$ and $T_{{}_{FSI}} (f)$ for ISI and FSI respectively, 
i.e.\ $T_{fi} \, = \, <\!f\,|\,T\,|\,i\!> \; \simeq \, T_{{}_{FSI}} (f) \; T^{\,(0)}_{fi} \; T_{{}_{ISI}} (i) $. 
Commonly used enhancement factors describe the elastic {\em onshell} scattering problem of the incoming or outgoing particles, so they don't have any reminder in the short ranged production process due to original time--reversal arguments of Watson. Yet from $\Delta E \, \Delta t \ge h\!\!\bar{} \;/2$ we know that the scattering system is going for a certain time of order $\Delta t$ offshell, while the amount of offshellness $\Delta E$ is strongly determined by the short range interaction process. It will be shown that the reminder of the enhancement factors in the threshold value of the short range production amplitude $T^{\,(0)}_{fi,thr}$ can be estimated by unitarity constraints on the T--matrix $T_{fi}$.
To test the validity of enhancement factors presently in use one has to check three minimal constraints: (a) enhancement factors have to be dimensionless, (b) for {\em no} ISI or FSI the respective enhancement factors have to be 1, (c) the S--matrix $S = 1 \, + \, i \, (2\pi )^4 \, \delta^{\, 4} (P_f - P_i\,) \, T$ has to be unitary.
Constraint (c) yields unitarity constraints on the T--matrix: 
assuming time--reversal--invariance ($<\!f\,|\,T\,|\,i\!> \stackrel{!}{=} <\!i\,|\,T\,|\,f\!>  = <\!f\,|\,T^{\,+}\,|\,i\!>^\ast$) I obtain after insertion of a complete set of relativistically normalized states \cite[p.\ 645 ff]{joa1}:
\[ \mbox{Im} \, <\!f\,|\,T\,|\,i\!> \, = \, \frac{1}{2} \; \sum\limits_{n} \; (2\pi )^4 \; \delta^{\, 4} (P_f - P_i\,) \;\;  <\!f\,|\,T\,|\,n\!> \; (<\!n\,|\,T\,|\,i\!>)^\ast \]
{\em In the approximation} that the contribution via Fock--states being different from the initial and final states is very small (which is not valid for no ISI {\em and} no FSI) one can formulate an approximate unitarity relation replacing the sum $\sum_{n}$ by $\sum_{n\,\in \{i,f\}}$.
Reexpressing the symbolic sum by integrals and applying Watson--Migdal factorisation with $T^{\,(0)}_{fi} \simeq T^{\,(0)}_{fi,thr}$ I arrive at the following unitarity constraint on the  enhancement factors for a particle production process $1+2\rightarrow 1^\prime + 2^\prime + \ldots + n^\prime$ close to threshold: 
\begin{eqnarray} 
\lefteqn{\mbox{Im} \, (T_{{}_{FSI}} (f) \; T^{\,(0)}_{fi,thr} \; T_{{}_{ISI}} (i) ) \; \simeq \frac{1}{2} \; \Big[ \,\, T^{\,\ast\,(0)}_{fi,thr} \; T^{\,\ast}_{{}_{ISI}} (i)}\nonumber \\
 & & 
\int\!
 \frac{d^3p_{1^\prime}}{(2\pi)^3\, 2\,\omega_{1^\prime}(|\vec{p}_{1^\prime}|)}  \cdot ... \cdot 
 \frac{d^3p_{n^\prime}}{(2\pi)^3\, 2\,\omega_{n^\prime}(|\vec{p}_{n^\prime}|)}  <\!f\,|\,T\,|\,1^\prime\ldots n^\prime\!>  T^{\,\ast}_{{}_{FSI}} (1^\prime\ldots n^\prime)  \nonumber \\
 & &\quad
+  \, T_{{}_{FSI}} (f) \; T^{\,(0)}_{fi,thr} \;
\int\!
 \frac{d^3p_1}{(2\pi)^3\, 2\,\omega_1(|\vec{p}_1|)} \,
 \frac{d^3p_2}{(2\pi)^3\, 2\,\omega_2(|\vec{p}_2|)} \nonumber \\
 & & \nonumber \\
 & & \quad\qquad\qquad\qquad\qquad\qquad\quad\;\;
 T_{{}_{ISI}} (12) \; (<\!12\,|\,T\,|\,i\!>)^\ast \, \Big] \; (2\pi )^4 \; \delta^{\, 4} (P_f - P_i\,) \nonumber 
\end{eqnarray}
In the limit of no ISI ($ T_{{}_{ISI}} (i)=1$, $<\!12\,|\,T\,|\,i\!>\simeq 0$) {\em or} no FSI ($T_{{}_{FSI}} (f)=1$, $<\!f\,|\,T\,|\,1^\prime\ldots n^\prime\!>\simeq 0$) it is just an integral constraint on the enhancement factors $T_{{}_{FSI}} (f)$ or $T_{{}_{ISI}} (i)$ respectively. 
\par As $T^{\,(0)}_{fi}\simeq const$ close to threshold, the energy dependence of the total cross section of a process with an $n$--particle final state close to threshold is mainly determined by a FSI-modified phasespace integral \cite{kle1,kle2}($s = P^2$):
\[ R^{\,FSI}_{\,n} (s)  = 
\int\!
 \frac{d^3p_{1^\prime}}{2\,\omega_{1^\prime}(|\vec{p}_{1^\prime}|)} \cdot \ldots \cdot 
 \frac{d^3p_{n^\prime}}{2\,\omega_{n^\prime}(|\vec{p}_{n^\prime}|)}
\, \delta^{\, 4} (p_{1^\prime} + \ldots + p_{n^\prime} - P) \,
{\left| T_{{}_{FSI}} (f)\right|}^2 
\] 
Without loss of generality I assume only FSI between particle $1^\prime$ and $2^\prime$. The enhancement factor now can be reexpressed either by the phaseshifts $\delta_\ell (\kappa)$ or Jost--functions 
$f_\ell (\kappa)\stackrel{!}{=}f^{\,\ast}_\ell (-\kappa^{\,\ast})$ of the $1^\prime2^\prime$--onshell--scattering problem ($\kappa := \sqrt{\lambda (s_{12}\,,m^2_{1^\prime},m^2_{2^\prime})}/(2(m_{1^\prime}+m_{2^\prime}))$)($\kappa^{2\ell+1}\cot \delta_\ell (\kappa) = -\,a^{-1}_\ell + O (\kappa^2)$)\cite{kle1}:
\[ T_{{}_{FSI}} (f) \simeq T_{{}_{FSI}} (1^{\,\prime}2^{\,\prime}) = \frac{\displaystyle \cot \delta_\ell (\kappa)- \frac{{\cal P}(\kappa)}{a_\ell \, \kappa^{2\ell+1}}}{\cot \delta_\ell (\kappa) - i} \stackrel{!}{=} \frac{\mbox{Re} f_\ell (\kappa)}{f_\ell (\kappa)} + \frac{{\cal P}(\kappa)}{a_\ell \, \kappa^{2\ell+1}} \frac{\mbox{Im} f_\ell (\kappa)}{f_\ell (\kappa)} \]
Here I used $(\cot \delta_\ell (\kappa) - i)^{-1}\stackrel{!}{=}(f_\ell (- \kappa) - f_\ell (\kappa))/(2i f_\ell (\kappa))$ (see e.g.\ \cite[p.\ 286]{joa1}).
${\cal P}(\kappa)$ is the offshell quantity defined in \cite{han1} carrying the reminder in the short ranged interaction. It either has to be calculated directly \cite{gar1,bar1} by a principle value integral or estimated from the unitarity constraint derived above. The limit $T_{{}_{FSI}} (1^{\,\prime}2^{\,\prime})\rightarrow 1$  for $\delta_\ell (\kappa)\rightarrow 0$ yields ${\cal P}(0) \rightarrow 0$ {\em for no FSI}.
As ${\cal P}(\kappa)$ depends on the nature of the short ranged interaction process, it has to be estimated for every single short ranged interaction diagram separately \cite{bar1}. 
Most authors upto now are using the approximations Re$f_\ell (\kappa)=0$ or $T_{{}_{FSI}} (1^{\,\prime}2^{\,\prime})=f^{-1}_\ell (\kappa)$ \cite{joa1,sib1} conflicting strongly with the unitarity constraint. A review on related models can be found in \cite{nis1}.
The only existing approach driven by unitarity is called Fermi--Watson Theorem (see e.g.\ \cite{joa1}). It states: $T_{fi} \, = \, \exp(i\,\mbox{Re}\, \delta_\ell (f)) \, |T_{fi}| \, \exp(i\,\mbox{Re} \, \delta_\ell (i))$. 
For ``well--behaved'' FSI--potentials $|T_{{}_{FSI}} (1^{\,\prime}2^{\,\prime})|^2$ can easily be Taylor--expanded in $\kappa$, i.e. $|T_{{}_{FSI}} (1^{\,\prime}2^{\,\prime})|^2 = \sum_\alpha\, c_{\,\alpha} \, \kappa^{\,\alpha}$. If the outgoing $1^\prime 2^\prime$--system is not bound, this Taylor--expansion is a threshold expansion of $R^{\,FSI}_{\,n} (s)$. E.g.\ for $n=3$ I obtain an expansion in $\eta := \eta_{\,12} := \sqrt{\lambda (s\,,m^2_{3^\prime},(m_{1^\prime}+m_{2^\prime})^2)/(4\, s \, m^2_{3^\prime})}$ ($2\bar{\mu} := (m_{1^\prime}+m_{2^\prime})/m_{3^\prime}$, $\Delta := \sqrt{1-(m_{1^\prime}-m_{2^\prime})^2/(m_{1^\prime}+m_{2^\prime})^2}$):
\begin{eqnarray} \lefteqn{R^{\,FSI}_{\,3} (s) =} \nonumber \\
 & = &  
\sum\limits_\alpha\, c_{\,\alpha} \left( \frac{\pi^2 \, m^{\,\alpha + 2}_{3^\prime}}{2^{\,\alpha +1} \, (2\bar{\mu})} \; \eta^{\,\alpha + 4} \!\int^{\,1}_{0}\!\!\!du \; \sqrt{u \; ( (1 + 2 \bar{\mu} ) \Delta^2 (1-u) )^{\alpha + 1}} + O(\eta^{\,\alpha + 6}) \right) \nonumber 
\end{eqnarray}
A similar expansion in $\eta_{\,23} := \sqrt{\lambda (s\,,m^2_{1^\prime},(m_{2^\prime}+m_{3^\prime})^2)/(4\, s \, m^2_{1^\prime})}$ (or $\eta_{\,31}$) can be performed for FSI between outgoing particles $2^\prime 3^\prime$ (or $1^\prime 3^\prime$). Close to threshold these expansions can be reformulated in terms of $\eta$ by ($m_{1^\prime}\ge m_{2^\prime}$):
\[ \left. 
\begin{array}{c}\eta_{\,23} \\ 
\eta_{\,31}
\end{array}
\right\} \; = \; \left\{ 
\begin{array}{c} (m_{3^\prime}/ m_{1^\prime} ) \\ 
(m_{3^\prime}/ m_{2^\prime} )
\end{array}
\right\} \cdot \frac{\eta}{2} \;\, \sqrt{2 \pm 2 \, \sqrt{1\,-\Delta^2} + (2\,\bar{\mu}) \, \Delta^2} \; + \; O(\eta^3)
\]
Of course the treatment of FSI as a sum of FSI between the various two--particle subsystems is not sufficent \cite{kle1}. A complete onshell--treatment would yield the the knowledge of the full outgoing elastic T--matrix. The offshell extension of this framework is non--trivial. Next to leading order terms have been attacked by \cite{moa1}. For corrections due to ISI and the flux factor see \cite{kle1}.
\par Because of essential singularities due to the penetration factor $C^{\,2}_0(\gamma )=2\pi\gamma \, (\exp(2\pi\gamma)-1)^{-1}$ in a Coulomb--Modified Effective Range Expansion \cite{aus1,hae1,hae2} with $\gamma := \alpha \, \sqrt{(\kappa^2+(m_{1^\prime} m_{2^\prime}/(m_{1^\prime}+m_{2^\prime}))^2)/\kappa^2} \; (\alpha \simeq 1/137)$ the Taylor--expansion described above can't be performed for a Coulomb potential $V(r)= 2\gamma\,\kappa/r $. Yet by determining $f_\ell (\kappa)$ for a regularised Coulomb potential $V(r)= 2\gamma\,\kappa \exp (-\mu \, r)/(r+\varepsilon ) $ using \cite{trl1} a regularised Modified Effective Range Function can be derived \cite{hae2}, which does not suffer from essential singularities mentioned. Now the Taylor--expansion of $|T_{{}_{FSI}} (1^{\,\prime}2^{\,\prime})|^2$ is well defined and the limits $\mu\rightarrow +0$ and $\varepsilon\rightarrow +0$ are straight forward.
\par An interesting test for FSI is the $\eta$--dependence of $\sigma (pp\rightarrow pp\pi^0)$ close to threshold. Experiments yield $\sigma (s)\propto \eta^{\,\alpha+4}$ with $\alpha \ge 0$ for $\eta < 0.15$ and $\alpha \simeq -2$ for $0.2 \le \eta \le 0.5$. As the $pp\pi^0$--system is not bound and theoretical models seem to provide no mechanism for $|T_{fi}|^2 \propto \kappa^{\,\alpha}$ with $\alpha < 0$, it seems that present models \cite{nis2} generate by hand inverse powers of $\kappa$ as it was done by \cite{kol1}, who replaced $(1+(a\,\kappa)^2)^{-1}$ by $(a\,\kappa)^{-2}$ in the Effective Range Factor. The real origin for $\alpha < 0$ still has to be explored. The ansatz $\sigma (pp\rightarrow pp\pi^0)\simeq D \, \eta \,(\zeta \eta)^3 (1+\sqrt{1+\zeta^2\eta^2}\,)^{-2}$ of \cite{fal1} has desired properties, if $\zeta$ is chosen such that $\zeta \eta \ll 1$ for $\eta < 0.15$ and $\zeta \eta \gg 1$ for $\eta \ge 0.2$. 

\end{document}